\magnification1200

\rightline{KCL-MTH-12-10}
%\rightline{hep-th/yymmnnn}

\vskip 2cm
\centerline
{\bf   Generalised BPS conditions}
\vskip 1cm
\centerline{ Peter West}
\centerline{Department of Mathematics}
\centerline{King's College, London WC2R 2LS, UK}
\vskip 2cm
%\centerline{and}
%\vskip 0.5cm
%\centerline{??,}
%\centerline {??,}
%\centerline{??}
\leftline{\sl Abstract}
We write down two $E_{11}$ invariant conditions which at low
levels reproduce the known half BPS conditions for type II theories.
These new conditions contain, in addition to the familiar
central charges, an infinite number of further charges which are required
in an underlying theory of strings and branes. We comment on the
application of this work to higher derivative string corrections. 
\vskip2cm
\noindent

\vskip .5cm

\vfill
\eject

The requirement  that  the solutions which occur in supersymmetric
theories  preserve a certain amount of the supersymmetry leads to
conditions on the charges that they carry. The most obvious  brane charges
are known  to occur as central charges in the supersymmetry algebra [1,2].
As a result, these   conditions can be found  by insisting that the
anti-commutator of the supercharges result in a matrix that has a
vanishing determinant, or more precisely,  has a certain number of
eigenvectors with zero eigenvalues.  The most familiar  BPS condition,
that is derived in this way,  relates the  space-time momentum to the
central charges.  Such conditions are known in quite a number of contexts
and in particular for black hole solutions they can be found in reference
[3].  
\par
In this paper we will consider the  BPS
conditions corresponding to  the preservation of half of the 
supersymmetries (half BPS conditions) in the type II
theories. These theories occur at low levels in the non-linear
realisation of $E_{11}$ [4,5,6,7,8,9] and the purpose of this paper is
to give an
$E_{11}$ formulation of these BPS conditions. The  brane
charges are contained in the first fundamental representation of $E_{11}$
associated with node one, denoted
$l_1$ [10,11,12,13,14]. Hence one would expect the BPS conditions to be
a condition on this representation. The $E_{11}$ Dynkin diagram is given
by 
\par
$$
\matrix{
&&&&&&& & & & &&& &\bullet &11&&&\cr 
&&&&&& & & & &&& & &| & && & \cr
\bullet &-&\bullet &-&\bullet&-&\bullet&-&\bullet&-&\bullet  &-
&\bullet&-&\bullet&-&\bullet&-&\bullet
\cr
1& &2&&3 &&4 &&5&&6& &7& &8& & 9&
&10\cr}
$$
\par
The central charges do not form  $U$-duality multiplet and must be
complimented by additional charges. By applying U-duality
transformations in lower dimensions to known charges, these additional
charges have been found sometime ago  for certain collections of
charges [15,16,17,18];  a review of this work can be found in [19]. The
$l_1$ representation contains the usual central charges but, when viewed
from the perspective of lower dimensions, the charges it contains 
automatically belong to U-duality multiplets.  It also provides a higher
dimensional origin for all the exotic charges required for this
completion. It follows that an $E_{11}$ formulation of the BPS conditions 
will be a generalisation of the usual BPS conditions, that is the ones
derived from the supersymmetry algebra,  as they must  also contain
the additional charges required by U-duality.  The
half BPS conditions are known to be quadratic
in the brane charges and as such  we  will look for
$E_{11}$ invariant conditions that are bilinear in the $l_1$ multiplet.
We will find two such conditions and show that when they are restricted to
contain  the familiar charges they  will agree with the known half BPS
conditions. 
\par
As seen from eleven dimensions the $l_1$ multiplet contains the objects
[10,11,13] 
$$
P_{\hat a} \ (0),\  Z^{\hat a_1\hat a_2}\ (1),
\  Z^{\hat a_1\ldots \hat a_5}\ (2)
,\  Z^{\hat a_1\ldots \hat a_7,b}\ (3),
\  Z^{\hat a_1\ldots \hat a_8}\ (3),
$$
$$
\  Z^{\hat b_1\hat b_2 \hat b_3, \hat a_1\ldots \hat a_8}\ (4),
\  Z^{(\hat c \hat d ), \hat a_1\ldots \hat a_9}\ (4),
\  Z^{\hat c\hat d,\hat a_1\ldots \hat a_9}\ (4),\ 
\  Z^{\hat c,\hat a_1\ldots \hat a_{10}}\ (4),\ 
Z\ (4)
$$
$$
Z^{\hat c, \hat d_1\ldots \hat d_4,\hat a_1\ldots \hat a_9}\ (5),\ 
Z^{\hat c_1\ldots \hat c_6,\hat a_1\ldots \hat a_8}\ (5),\ 
Z^{\hat c_1\ldots \hat c_5,\hat a_1\ldots \hat a_9}\ (5),\ 
$$
$$
Z^{\hat d_1,\hat c_1 \hat c_2 \hat c_3,\hat a_1\ldots \hat a_{10}}\
(5),\  Z^{\hat c_1 \ldots \hat c_4,\hat a_1\ldots \hat a_{10}}\
(5,-2),\  Z^{\hat a_1\ldots \hat a_{11}, (\hat c_1\hat c_2,\hat c_3 )}\
(5),\  Z^{\hat a_1\ldots \hat a_{11},\hat c,\hat b_1\hat b_2}\  (5),\ 
\ldots 
\eqno(1)$$
The indices take the values $\hat a_1, \hat b_1, \dots = 1,2,\ldots , 11$.
These generators satisfy irreducibility conditions  such as
$Z^{[\hat a_1\ldots\hat a_7,\hat  b]}=0, \ldots $. The blocks of eleven
indices are inert under SL(11) transformations. As with all representations of
$E_{11}$ the states of the $l_1$ representation can be classified by  a
level and the number in brackets gives the level relevant to the eleven
dimensional theory. We note that  a
level $n$ generator has
$3n-1$ indices.  The
eleven-dimensional theory emerges from the $E_{11}$ non-linear realisation
if one decomposes $E_{11}$ in terms of its SL(11) subalgebra which is the
one that emerges if one deletes node eleven in the above $E_{11}$ Dynkin
diagram. 
\par
In any dimension the $l_1$ representation has as its lowest component the
space-time momenta $P_a$, $a=1,\ldots ,d$, while the form charges in $d$
dimensions   are given in the table [12,13,20] below 

\bigskip

$$\halign{\centerline{#} \cr
\vbox{\offinterlineskip
\halign{\strut \vrule \quad \hfil # \hfil\quad &\vrule Ê\quad \hfil #
\hfil\quad &\vrule \hfil # \hfil
&\vrule \hfil # \hfil Ê&\vrule \hfil # \hfil &\vrule \hfil # \hfil &
\vrule \hfil # \hfil &\vrule \hfil # \hfil &\vrule \hfil # \hfil &
\vrule \hfil # \hfil &\vrule#
\cr
\noalign{\hrule}
d&G&$Z$&$Z^{a}$&$Z^{a_1a_2}$&$Z^{a_1\ldots a_{3}}$&$Z^{a_1\ldots a_
{4}}$&$Z^{a_1\ldots a_{5}}$&$Z^{a_1\ldots a_6}$&$Z^{a_1\ldots a_7}$&\cr
\noalign{\hrule}
8&$SL(3)\otimes SL(2)$&$\bf (3,2)$&$\bf (\bar 3,1)$&$\bf (1,2)$&$\bf
(3,1)$&$\bf (\bar 3,2)$&$\bf (1,3)$&$\bf (3,2)$&$\bf (6,1)$&\cr
&&&&&&&$\bf (8,1)$&$\bf (6,2)$&$\bf (18,1)$&\cr Ê&&&&&&&$\bf (1,1)$&&$
\bf
(3,1)$&\cr Ê&&&&&&&&&$\bf (6,1)$&\cr
&&&&&&&&&$\bf (3,3)$&\cr
\noalign{\hrule}
7&$SL(5)$&$\bf 10$&$\bf\bar 5$&$\bf 5$&$\bf \bar {10}$&$\bf 24$&$\bf
40$&$\bf 70$&-&\cr Ê&&&&&&$\bf 1$&$\bf 15$&$\bf 50$&-&\cr
&&&&&&&$\bf 10$&$\bf 45$&-&\cr
&&&&&&&&$\bf 5$&-&\cr
\noalign{\hrule}
6&$SO(5,5)$&$\bf \bar {16}$&$\bf 10$&$\bf 16$&$\bf 45$&$\bf \bar
{144}$&$\bf 320$&-&-&\cr &&&&&$\bf 1$&$\bf 16$&$\bf 126$&-&-&\cr
&&&&&&&$\bf 120$&-&-&\cr
\noalign{\hrule}
5&$E_6$&$\bf\bar { 27}$&$\bf 27$&$\bf 78$&$\bf \bar {351}$&$\bf
1728$&-&-&-&\cr Ê&&&&$\bf 1$&$\bf \bar {27}$&$\bf 351$&-&-&-&\cr
&&&&&&$\bf 27$&-&-&-&\cr
\noalign{\hrule}
4&$E_7$&$\bf 56$&$\bf 133$&$\bf 912$&$\bf 8645$&-&-&-&-&\cr
&&&$\bf 1$&$\bf 56$&$\bf 1539$&-&-&-&-&\cr
&&&&&$\bf 133$&-&-&-&-&\cr
&&&&&$\bf 1$&-&-&-&-&\cr
\noalign{\hrule}
3&$E_8$&$\bf 248$&$\bf 3875$&$\bf 147250$&-&-&-&-&-&\cr
&&$\bf1$&$\bf248$&$\bf 30380$&-&-&-&-&-&\cr
&&&$\bf 1$&$\bf 3875$&-&-&-&-&-&\cr
&&&&$\bf 248$&-&-&-&-&-&\cr
&&&&$\bf 1$&-&-&-&-&-&\cr
\noalign{\hrule}
}}\cr}$$
\par
The definition of the level depends on the dimension one is in; the
level of the charges in $d$ dimensions is the number of
upper $d$-dimensional indices they carry.
\par
We begin by finding an  $E_{11}$ formulation of the most common  BPS
condition which is  of the generic form $p^2+Z^2=0$ where $p_\mu$ is the
space-time momentum and $Z$ one of the central charges. Since this is
bilinear in the charges and consists of a single  equation we are looking
for an $E_{11}$ invariant that is bilinear in the
$l_1$ representation. Let us denote the charges in the $l_1$
representation by $l$. Their transformation under an
$E_{11}$ group element
$g_0$  can be  written in the form 
$$
l^\prime = D(g_0^{-1}) l ,\quad {\rm or \ in \ components}\quad 
l^\prime_N=  D(g_0^{-1})_N{}^M l_M
\eqno(2)$$
where $D$ are  the matrices of the $l_1$  representation and $l$ is taken
to be a column vector. 
\par
The
non-linear realisation of
$E_{11}$ is constructed from an $E_{11}$ group element $g(\xi)$ where
$\xi$ parameterise the group element. In the non-linear realisation the
$\xi$ become  the fields of the theory. This group element is subject to
the transformations  
$$
g(\xi)\to g (\xi^\prime)= g_0 g(\xi), \ \ \ g_0\in E_{11},\ \ {\rm as \ 
well \  as} \
\ \ g(\xi)\to g(\xi)h, \ \ \ h\in I(E_{11})
\eqno(3)$$
The group element $g_0\in E_{11}$ is a rigid transformation, that is, a 
constant,  while $h\in I(E_{11})$ is a local transformation.
Here $I$ is the Cartan involution and $I(E_{11})$ is the Cartan
involution invariant subgroup of $E_{11}$ and as a result $I(h)=h$.  For
a discussion of the Cartan involution and
$I(E_{11})$ see  section 16.7.3 of [21],  or earlier
papers on $E_{11}$. We note that the action of the Cartan involution
preserves the order of  two group elements
$g_1$ and
$g_2$,  that is, $I(g_1g_2)=g_1g_2$. 
\par
Using the element $g(\xi)$ that occurs in the non-linear realisation we
can consider the states 
$$
L(\xi)\equiv D(g(\xi)^{-1}) l
\eqno(4)$$
It is straightforward to verify that they transform under rigid and local
transformation as follows 
$$
L^\prime = D(g(\xi)^{-1})D(g_0^{-1})l= D(g(\xi^\prime)^{-1})l=
L(\xi^\prime),\quad {\rm and} \quad L^\prime = D(h^{-1}) L(\xi)
\eqno(5)$$
In the maximal supergravity theories 
the supercharges do not transform under the U duality group but under its
Cartan involution invariant subgroup and as a result  the central charges
also transform under this latter group. As such we can expect the BPS
conditions to be constructed from  $L$. 
\par
From the $l_1$ representation  one can construct what is
called the dual twisted representation $l_{ID}$,  
which transforms  as 
$$
l_{ID}^\prime = l_{ID} D(I(g_0))  ,\quad {\rm or \ in \ components}\quad 
l_{ID}{}^{\prime M}= l_{ID}{}^{N} D(g_0^{-1})_N{}^M 
\eqno(6)$$
where $l_{ID}$ is taken to be a row vector. Clearly given the matrices
$D$ one can define such a transformation and it is trivial to verify that
it is indeed a representation.  For a more detailed discussion of such
representations and some of the other steps above see the appendix B of
reference [22].  In fact the dual twisted representation $l_{ID}$ has the
same highest weight as the original representation $l$ as the Cartan
subalgebra elements are preserved by the combined action of an inverse
transformation and the Cartan involution. As such the representation
$l_{ID}$  can be identified with the transpose of the original
representation, that is, with 
$l^T$.  The analog of $L$ for the twisted dual representation is
$L_{ID}(\xi)=l^T D((Ig(\xi)))$
\par
Using equation (3) it is trivial to verify that 
$$
{ L}^2
\equiv l_{ID} D(I(g(\xi)))D(g(\xi)^{-1}) l=l^T D(I(g(\xi)))D(g(\xi)^{-1})
l=  l^T D(M(\xi)^{-1}) l= L^T(\xi) L(\xi)
\eqno(7)$$
is invariant under both the rigid and local
transformations of equation, that is, the transformations of the
non-linear realisation (3). In the second step  we have defined the
object  
$M(\xi)\equiv g(\xi)I(g(\xi)^{-1}$. 
\par
It will be useful to rewrite the above expression for ${ L}^2$. When
constructing  the non-linear realisation of the semi-direct 
product of $E_{11}$ and
the generators of the $l_1$ representation, denoted by 
$E_{11}\otimes_s l_1$,  one automatically 
finds a generalised space-time, whose elements $x^N$ are in one to one
correspondence with the $l_1$ representation,   that is 
equipped with a generalised vielbein [10,7]. We will not give an account
of this construction here but we  will  recall the parts we need. For each
element of the
$l_1$ representation we introduce  corresponding generators, that  we
denote by  $\tilde l$, and which  by definition have commutators with the
generators of the 
$E_{11}$ algebra such that  for any $g_0\in E_{11}$  transformation 
$$
g_0^{-1}\tilde l g_0= D(g_0)  \tilde l ,\quad {\rm or \ in \
components}\quad  g_0^{-1}  \tilde l_N g_0=  D(g_0)_N{}^M \tilde l_M
\eqno(8)$$
The infinitesimal version of this relation are just the commutators of
the $E_{11}$ generators and the $\tilde l$ generators. These are given in
the appendix. It turns out that the generalised vielbein
$E_N{}^A$   is
given by 
$$
g^{-1}(\xi ) dx\cdot \tilde  l g (\xi ) = D(g(\xi ) ) \tilde l= dx \cdot
E\cdot 
\tilde l  
$$
or in components
$$\quad  g^{-1}(\xi ) dx\cdot  \tilde  l g(\xi ) = dx ^N
D(g)_N{}^A\tilde  l_A= dx^N E_N{}^A\tilde  l_A
\eqno(9)$$
The $x^N$ are the coordinates of the generalised space-time that arises 
in the non-linear realisation of  $E_{11}\otimes_s
l_1$,  but in this paper they  will just act as dummy
variables.  Equation (9) contains the expression that  arises in the
non-linear realisation and defines the generalised vielbein $E_N{}^A$. 
 We recogonise, using matrix notation,  that 
$$
E= D(g(\xi)) , \quad {\rm and \ so }\quad  E^{-1}= D(g(\xi)^{-1})
\eqno(10)$$
Using the known $E_{11}$ commutators with the generators
in the
$l_1$ representations, given in the appendix,  it is straightforward to
compute the generalised vielbein in any dimension, at least at low levels
and below we will give an example. 
\par
Substituting  equation (9) into equation (7) we find that 
$$
L^2\equiv l^T (E^{-1})^TE^{-1} l= L_A L_A
\eqno(11)$$
where as previously defined $L_A= (E^{-1}) _A{}^N l_N$. Technically the
transpose acting on a group element is defined by $g^T= I(g^{-1})$, but
the corresponding action on the matrix representative is indeed the
transpose. 
\par
We take half BPS states to obey the $E_{11}$ invariant condition 
$$
L^2=0 
\eqno(12)$$
\par
It has been found that the half BPS states also obey another
condition, or rather a U-duality multiplet of conditions, which are also
bilinear in the charges [15,16,17,18,19]. As such the $E_{11}$ formulation
of this condition should also be bilinear in the
$l_1$ representation. Guided by the fact that the first component in the
U-duality multiplet is an object with one upper space-time vector index
 we take this condition to be that the tensor product of two
$l_1$ representations  restricted to the fundamental representation
associated with node ten, that is 
$l_{10}$, vanishes.  In symbols this is the condition is given by 
$$
l_1\otimes l_1 |_{l_{10}}=0
\eqno(13)$$
Clearly this condition is $E_{11}$ invariant. We note that it 
only involves the brane charges and not the generalised vielbein and so it
does not involve the fields of the theory. When viewed from eleven
dimensions the
$l_{10}$ representation contains the following objects 
$$
S^{\hat a} (1,1); S^{\hat a_1\ldots \hat a_4} (2,1), S^{\hat a_1\ldots
\hat a_7} (3,1); S^{\hat a_1\ldots
\hat a_6,\hat b} (3,1); 
$$
$$
S^{\hat a_1\ldots \hat a_{10}} (4,1), 
S^{\hat a_1\ldots \hat a_{9}, \hat b} (4,2), 
S^{\hat a_1\ldots \hat a_{8}, \hat b_1 \hat b_2} (4,1) , S^{\hat a_1\ldots
\hat a_{8}, (\hat b_1 \hat b_2)} (4,1),  S^{\hat a_1\ldots \hat a_{7},
\hat b_1 \hat b_2 \hat b_3} (4,1), 
$$
$$
S^{\hat a_1\ldots \hat a_{11}, \hat b_1\hat b_2} (5,3), S^{\hat a_1\ldots
\hat a_{11}, (\hat b_1\hat b_2)} (5,3), 
$$
$$
S^{\hat a_1\ldots \hat a_{10}, \hat b_1 \hat b_2\hat b_3} (5, 3),
 S^{\hat a_1\ldots \hat a_{10}, \hat b_1
\hat b_2, \hat c_1} (5,3), 
S^{\hat a_1\ldots \hat a_{10}, (\hat b_1 \hat b_2\hat b_3)} (5,1),
$$
$$
S^{\hat a_1\ldots \hat a_{9}, \hat b_1 \hat b_2\hat b_3\hat b_4} (5, 2),
S^{\hat a_1\ldots \hat a_{9}, \hat b_1 \hat b_2\hat b_3, \hat c_1} (5, 2),
$$
$$
S^{\hat a_1\ldots \hat a_{8}, \hat b_1 \ldots \hat b_5} (5,1),
S^{\hat a_1\ldots \hat a_{8}, \hat b_1 \ldots \hat b_4,\hat c} (5,1),
S^{\hat a_1\ldots a_{7}, \hat b_1 \ldots \hat b_6} (5,1),\dots 
\eqno(14)$$
The indices grouped together are anti-symmetrised except where indicated
otherwise and the obey the usual irreduciblility constraints, for
example $S^{[\hat a_1\ldots \hat a_6,\hat b]}=0$. One can find these
results analytically using the techniques of references [23,24] which are
reviewed in section 16.6.2 reference [21], or 
by  using the Nutma computer  programme [25]. The numbers in brackets are
the level,  appropriate to the eleven dimensional decomposition, and the
multiplicity respectively. We have taken the highest weight, that is the
first,  state to have level one. 
\par
To evaluate the components of equation (13) we must construct the
$l_{10}$ representation from two $l_1$ representations. The first step is
to write down objects with the correct SL(11) index structure
and level from two $l_1$ representations with arbitrary coefficients. The
coefficients can be determined by starting with the first component,
which is given by 
$S^{\hat a}= P_{\hat b} Z^{\hat b\hat a}$,  and varying it under the
$E_{11}$ transformation 
$\Lambda _{\hat c_1\hat c_2\hat c_3}R^{\hat c_1\hat c_2\hat c_3}+\Lambda
^{\hat c_1\hat c_2\hat c_3}R_{\hat c_1\hat c_2\hat c_3}$.  One finds, using the
commutators in the appendix,  that 
$$
\delta S^{\hat a}= [\Lambda _{\hat c_1\hat c_2\hat c_3}
R^{\hat c_1\hat c_2\hat c_3}+\Lambda ^{\hat c_1\hat c_2\hat c_3}R_{\hat
c_1\hat c_2\hat c_3}, P_{\hat b} Z^{\hat b\hat a}] 
$$
$$
=- \Lambda_{\hat c_1\hat c_2\hat c_3}( Z^{\hat c_1\hat c_2\hat c_3
\hat a\hat b}P_{\hat b}+ 3Z^{\hat a\hat c_1}Z^{\hat c_2\hat c_3})   
= \Lambda_{\hat c_1\hat c_2\hat c_3}
S^{\hat a\hat c_1\hat c_2\hat c_3 } 
\eqno(14)$$
and hence the expression for $S^{a_1\ldots a_4}$ given below. We note
that $Z^{\hat a_1[\hat a_2}Z^{\hat a_3\hat a_4]}
= Z^{[\hat a_1\hat a_2}Z^{\hat a_3\hat a_4]}$. At the next level 
we have 
$$
\delta S^{\hat a_1\hat a_2\hat a_3a_4}= [\Lambda
_{\hat c_1\hat c_2\hat c_3}R^{\hat c_1\hat c_2\hat c_3}, 
S^{\hat a_1\ldots \hat a_4} ]=
\Lambda_{\hat c_1\hat c_2\hat c_3}( 
-6Z^{[\hat a_1\hat a_2|\hat c_1\hat c_2\hat c_3}Z^{|\hat a_3\hat a_4]}
+3Z^{\hat c_1\hat c_2}Z^{\hat c_3 \hat a_1\hat a_2\hat a_3\hat a_4}
$$
$$
+ P_{\hat b} Z^{b\hat a_1\hat a_2\hat a_3\hat a_4\hat c_1
\hat c_2,\hat c_3}
+ P_{\hat b}Z^{b\hat a_1\hat a_2\hat a_3\hat a_4\hat c_1
\hat c_2\hat c_3} )=
\Lambda_{\hat c_1\hat c_2\hat c_3}S^{\hat a_1\hat a_2\hat a_3\hat a_4
\hat c_1\hat c_2,\hat c_3}+
\Lambda_{\hat c_1\hat c_2\hat c_3}S^{\hat a_1\hat a_2\hat a_3\hat a_4
\hat c_1\hat c_2\hat c_3}
\eqno(15)$$
Taking  anti-symmetry in all the indices it is straightforward to
read off the expression for $S^{a_1\ldots a_6 a_7} $ given below. 
The expression for $S^{a_1\ldots a_6,b} $ can also be found after 
 some non-trivial manipulations. 
We have found that 
$$
S^{\hat a} =P_{\hat b} Z^{\hat b\hat a} ; S^{\hat a_1\ldots \hat a_4}
 = -3Z^{[\hat a_1\hat a_2}Z^{\hat a_3\hat a_4]}
+P_{\hat b}Z^{\hat b\hat a_1\hat a_2\hat a_3\hat a_4}, 
$$
$$ S^{\hat a_1\ldots \hat a_7}  = -3Z^{[\hat a_1\hat a_2} 
Z^{\hat a_3\ldots \hat a_7]}
 +P_{\hat b}Z^{\hat b\hat a_1\ldots \hat a_7}, 
$$
$$
S^{\hat a_1\ldots \hat a_6,\hat b} =P_{\hat c} 
Z^{\hat c\hat a_1\ldots \hat a_6, \hat b} 
+{3.6.5\over 7}( Z^{\hat b[\hat a_1}Z^{\hat a_2\ldots \hat a_6]}
-Z^{[\hat a_1\hat a_2}Z^{\hat a_3\ldots
\hat a_6]\hat b}), \ldots 
\eqno(17)$$
The half BPS condition of equation (13) takes the form 
$$
S^a=0=S^{\hat a_1\ldots \hat a_4}= S^{\hat a_1\ldots \hat a_7}  =S^{\hat
a_1\ldots \hat a_6,\hat b}=\ldots 
\eqno(18)$$
We have used the symbol $S$ to denote both the components of the abstract
$l_{10}$ representation and the $l_{10}$ constructed from two $l_1$
representations. 
Although we have given the
$l_1$ and 
$l_{10}$ multiplets in eleven dimensions one can carry out the usual
dimensional reduction process to find the corresponding results in  lower
dimensions. 
\par
We will now evaluate the above half BPS conditions for the type II theory
in
$d$ dimensions at level zero.  The $d$ dimensional theory emerges  
from  the non-linear realisation  if we  decomposing $E_{11}$ into the
algebra $GL(d)\otimes E_{11-d}$. The  Dynkin diagram of this algebra 
remains when we  delete node d in the
$E_{11}$ Dynkin diagram and the objects that occur in  this decomposition
can be classified by a level which is associated with node $d$. The first
factor, that is
$GL(d)$,  corresponds to
$d$ dimensional gravity and the second fact is the U-duality symmetry in
$d$ dimensions. At level zero we have  the  graviton in $d$ dimensions
and the scalar fields that belong to the non-linear realisation of
$E_{11-d}$ while in the $l_1$ representation we have the
space-time momentum
$p_\mu$  in $d$ dimensions and the charges, denoted $Z$,  that are scalars
with respect to the GL(d) symmetry, but transform with respect to the
$E_{11-d}$ in the representations listed in the first column of table
one.  The generalised  vielbein at level zero is of the form 
$$
E=(\det e )^{-{1\over 2}} \left(\matrix {e_\mu{}^a&0\cr 
0&{\cal E}^{(0)}\cr}\right)
 \eqno(19)$$
where $e_\mu{}^a$ is the vielbein in the $d$ dimensional space-time
and ${\cal E}^{(0)}$ is the vielbein in the internal sector at level zero,
with the factor of 
$(\det e )^{-{1\over 2}} $ taken out.  The factor $(\det e )^{-{1\over 2}}
$ drops out of the BPS condition; it has its origins in the
unconventional commutator between the generators of GL(d) and the
space-time translations which follows from $E_{11}$.   As a result the 
 half BPS condition of equation (12) becomes 
$$
 p^2+ Z^T ({\cal E}^{(0)T})^{-1}({\cal E}^{(0)})^{-1} Z=0
\eqno(20)$$
\par
The condition of equation (1.13) also simplifies if we are at level zero.
For dimensions $4\le d\le 10$ we can set all objects with  any block 
of anti-symmetrised indices that contains more than seven indices to zero
and equation (17) now takes the form 
$$
P_iZ^{ij}=0, \quad -3Z^{[i_1i_2}Z^{i_3i_4]}+ P_jZ^{ji_1i_2i_3i_4}=0 ,
\quad  -3Z^{[i_1i_2}Z^{i_3\ldots i_7]}+P_k Z^{i_1\ldots i_7, k} =0
$$
$$ 
P_k Z^{ki_1\ldots i_6, j}+{3.6.5\over 7}(
Z^{j[i_1}Z^{i_2\ldots i_6 ]}- Z^{[i_1i_2}Z^{i_3\ldots i_6]j})=0,
\ldots
\eqno(21)$$ 
These agree with the results given in [19].
\par
It is instructive to evaluate the half BPS conditions of equations
(20)  and (21)  explicitly  for seven dimensions where $E_4=SL(5)$.
Apart from the graviton in seven dimensions we have fourteen scalars which
belong to the non-linear realisation of SL(5) with local subgroup SO(5).
The latter  arise from the eleven dimensional fields as $h_i{}^j$ and
$A_{i_1i_2i_3}$ where
$i,j,\ldots =1,2,3,4$. As such the part of the $E_{11}$ group element
$g(\xi)$ in the internal sector and at level zero is given by 
$$
g^{(0)}(\xi)= e^{h_i{}^j K^i{}_j}e^{{1\over
3!} A_{i_1i_2i_3}R^{i_1i_2i_3}}
\eqno(22)$$
The superscript 0 just indicates that we are at level zero.
The charges $Z$ of the 
$l_1$ representation at level zero transform as the 10 of SL(5) and
arise from eleven dimensions as
$Z=\{P_i, Z^{i_1i_2}\}$. As a result in the internal sector and at level
zero 
$$
(dx\cdot l)^{(0)}= dx^iP_i+ dx_{i_1i_2}Z^{i_1i_2}
\eqno(23)$$
It is almost trivial, using equation (22) and the commutators of the
appendix,  to verify that the generalised vielbein in the internal sector
is given by 
$$
{\cal E}^{(0)}= \left(\matrix{
e_{\bar i}{}^j& -{1\over 2}e_{\bar i} {}^k C_{k i_ii_2} \cr
0&(e^{-1})_{i_1i_2}{}^{{\bar j}_1{\bar j}_2}\cr}\right),\quad {\rm and }
\quad 
({\cal E}^{(0)})^{-1}= \left(\matrix{
(e^{-1})_i{}^{\bar j}& +{1\over  2} C_{i k_ik_2} e_{\bar j_1\bar
j_2}{}^{k_1k_2}\cr 0&e_{{\bar j}_1{\bar j}_2}{}^{i_1i_2}
\cr}\right),
\eqno(24)$$
where $e_{\bar i}{}^i=(e^h)_{\bar i}{}^i$,  
$e_{{\bar j}_1{\bar j}_2}{}^{i_1i_2}= e_{{\bar j}_1} {}^{[i_1} e_{\bar
j_2} {}^{i_2] }$ and $e^{-1}{}_{{ i}_1{i}_2}{}^{\bar j_1\bar j_2}
=(e^{-1}){}_{{ i}_1}^{[\bar j_1}(e^{-1}){}_{{ i}_2}{}^{\bar j_2]}$. We are
using
${\bar i},\bar j,\ldots $ as world indices in the four dimensional
internal space. The construction of the non-linear realisation in
seven dimensions at level zero in the internal sector was given in detail
in  reference [26]. This reference also contains the generalised
vielbein in dimensions $4\le d\le 7$. The charges   referred to the
tangent space, denoted previously by $L$,  are then given by 
$$
L = ({\cal E}^{(0)})^{-1} Z=  \left(\matrix{
(e^{-1})_i{}^{\bar j} P_{\bar j} +{1\over 2} C_{i k_ik_2} e_{\bar j_1\bar
j_2}{}^{k_1k_2}
Z^{\bar j_1\bar j_2}\cr e_{{\bar j}_1{\bar j}_2}{}^{i_1i_2}Z^{\bar
j_1\bar j_2}
\cr}\right)
\eqno(25)$$
which can be substituted into the half BPS condition of equation (20). 
\par
For the seven dimensional theory the half BPS condition of equation (21)
at level zero takes the form 
$$
P_iZ^{ij}=0
\eqno(26)$$
\par
We now consider the half BPS condition when we keep the charges at the
next level in the
$l_1$ representation where  we find a charge with a
$d$-dimensional vector index which belongs to the 
representations of $E_{11-d}$ given in the second column of the table.
These are the charges carried by  the strings in the $d$-dimensional
theory and we will denote them by the column vector
$Z^a$.  Keeping the
$E_{11}$ fields still only at level zero the generalised vielbein takes
the from 
$$
E=(\det e )^{-{1\over 2}} \left(\matrix {e_\mu{}^a&0&0\cr 
0&{\cal E}^{(0)}&0\cr
0&0&{\cal E}^{(1)}\cr}\right)
 \eqno(27)$$
and so the half BPS condition of equation (12) takes the form  
$$
 p^2+ Z^T ({\cal E}^{(0)T})^{-1}({\cal E}^{(0)})^{-1} Z+
Z^{aT} ({\cal E}^{(1)T})^{-1}({\cal E}^{(1)})^{-1} Z_a=0
\eqno(28)$$
\par
It is instructive to explicitly evaluate this condition for the seven
dimensional theory. The charges at level one  belong to the $\bar 5$
representations of SL(5) and as seen from the view point of 
eleven dimensions they arise as 
$Z^a=\{ Z^{ai}, Z^{ai_1\ldots i_4}\}$ where $i,i_1,\ldots =1,2,3,4$. To
compute the generalised vielbein at level one we consider 
$$
(dx\cdot l)^{(1)}= dx_{ai} Z^{ai}+ dx_{a i_1i_2}Z^{ai_1i_2}
\eqno(29)$$
Using equation (9) and the commutators of the
appendix,  we find  that the generalised vielbein at level one in the
internal sector is given by 
$$
{\cal E}^{(1)}= \left(\matrix{
(e^{-1})_j{}^{\bar i}& -{1\over 3!}(e^{-1})_{[j_1}{}^{\bar i}  A_{j_2
j_2j_3]}
\cr 0&(e^{-1})_{j_1j_2j_3j_4}{}^{{\bar i}_1{\bar
i}_2\bar i_3\bar i_4}\cr}\right),
\eqno(30)$$
with the inverse 
$$
({\cal E}^{(1)})^{-1}= \left(\matrix{
e_{\bar k}{}^j & {1\over 3!} A_{ m_2m_3m_4} e_{\bar k_1\bar
k_2\bar k_3\bar k_4}{}^{j m_2m_3m_4}\cr 0&e_{\bar k _1\dots
\bar k_4}{}^{{j}_1\ldots { j}_4}
\cr}\right),
\eqno(31)$$
Substituting into equation (28) one finds the BPS condition. For
situations in which the point particle charges vanish  one
finds a constraint for just the string charges. 
\par
The generalisations to include higher levels in the $l_1$ representation,
and indeed higher levels in the $E_{11}$ fields,  is straightforward and
the half BPS condition can be readily computed,  at least at low levels.  
\par
We will now evaluate the $E_{11}$ half BPS condition of equations (12)
and (13) in the familiar context  of 
 ten dimensions at level zero.  The  non-linear realisation of
$E_{11}\otimes _s l_1$, first advocated in [10],   was carried out in
in this context in reference [27]. It is just contains the fields of the
IIA NS-NS sector but they depend on the coordinates
$x^a$ and $y_a$ which transform as  the vector representation of
O(10,10). It is nothing but the so called doubled field theory [28]. The
level one non-linear realisation contains the fields in the R-R sector
[29].  To find the IIA theory in ten dimensions one
deletes node ten in the
$E_{11}$ Dynkin diagram to find the Dynkin diagram of the
algebra SO(10,10) and so one must decompose $E_{11}$ into representations
of SO(10,10). The non-linear realisation of
$E_{11}$ at lowest level  contains the fields of the NS-NS sector, namely
$h_a{}^b$,
$A_{ab}$ and
$a$, while the
$l_1$ representation contains the charges
$P_a$ and
$P^{\bar a}$ which form the vector representation of O(10,10). The
corresponding non-linear realisation, and so the dynamics, was worked out
in detail at level zero  in [27] and level one in [29]. In the first paper
we denoted the level zero generators in the $l_1$ representation by 
the symbols $P_a$ and $Q^a$ while in the second paper, and in this paper, 
we use the same
$P_a$ but $P^{\bar a}={1\over 2} Q^a$. In terms of the 
eleven-dimensional charges $Z^{a11}= -P^{\bar a}$. 
\par
The generalised vielbein, at lowest level,  was found [27, 29] to be given
by 
$$
 E_N{}^A = (\det e )^{-{1\over 2}}\left(\matrix {e&
eAe^{-{1\over 2}a}\cr 0&e^{-1T}e^{-{1\over 2}a}\cr}\right)
\eqno(32)$$
and so the charges referred to the tangent space take the form 
$$
L_A= (E^{-1})_A{}^N l_N= (\det e )^{{1\over 2}}\left(\matrix {
(e^{-1})_a{}^\mu P_\mu -e^{a\over 2}A_{a\nu}P^{\bar \nu}\cr
e^{a\over 2}e_\mu{}^aP^{\bar \mu} \cr}\right)
\eqno(33)$$
As such the half BPS condition of equation (12) becomes 
$$
L^2= L_A L_A= \left(\matrix {P_\mu ,P^{\bar
\mu}\cr}\right)M^{-1} \left(\matrix {P_\mu \cr
P^{\bar\mu}\cr}\right)=0
\eqno(34)$$
where $M=EE^T$. The other half BPS constraint of equation (13) takes the
simple form 
$$
P_\mu P^{\bar \mu}=0
\eqno(35)$$
\par
If we consider the ten dimensional  string to be compactified on a torus 
 then the Kaluza-Klein momenta and winding modes are identified with
$P_i$ and
$P^{\bar i}$ respectively. As a result, suppressing the radii of the
tori,  we must set these equal to the integers $m$ and $n$ respectively.
In this case the half BPS condition of equation (34) becomes
by 
$$
p^2+ \left(\matrix {m ,n\cr}\right)M^{-1}
\left(\matrix {m \cr n\cr}\right)=0
\eqno(36)$$
where $p_\mu$ is the momentum in the uncompactified space-time and the
second condition of equation (35)   takes the form 
$$ m\cdot n=0
\eqno(37)$$
We recognise equations (36) and (37) as the well known half BPS
conditions for the compactified string. 
\par
We now consider the conditions if only a quarter of the original
supersymmetry survives from and $E_{11}$ viewpoint. Examining the
supersymmetry algebra one finds that these conditions are quartic in the
central charges, see [19] for a very  telegraphed account. Let us define 
$$
s_{10}\equiv l\otimes l|_{l_{10}}
\eqno(38)$$
The quarter BPS condition should revert to the half BPS
conditions if
$s_{10}=0$ and so we propose that quarter BPS states should satisfy 
$$
(l^T D(M^{-1}) l)^2= s_{10}^T D(M^{-1})s_{10}
\eqno(39)$$
Now the matrix $D$ is that appropriate to the representation on which it
acts. 
Guided by an analysis of the supersymmetry algebra one finds that if
$s_{10}$ does not vanishes then demanding a quarter supersymmetry leads
to  the constraints 
$$
S^{[a_1}S^{a_2\ldots a_5]}=0, \ S^{[ a_1\ldots a_4} S^{a_1\ldots a_4
]}=0\ , \ldots 
\eqno(40)$$
As such one should also insist on the condition 
$$
s_{10}\otimes s_{10}|_{l_6}=0
\eqno(41)$$
that is, the tensor product of the two $s_{10}$ representations, each of
which is  constructed from the two $l_1$ representations, when projected
in the first fundamental representation associated with node six, denoted
$l_6$, vanishes. 
It is straightforward to evaluate these conditions at lowest level. 
\medskip
In this paper we have proposed two $E_{11}$ conditions that when
evaluated at lowest level agree with the known  half BPS conditions. 
We have not derived these conditions from first principles, that is
supersymmetry preservation. However, they are the unique
$E_{11}$ invariant conditions which are bilinear in the brane charges and
in the case of the second condition of equation (13) also have a first
component that is an object with one upper SL(11) index. The conditions
are very simple and easy to evaluate. These  $E_{11}$
conditions contain an infinite number of fields and charges in addition to
those with which people are familiar, however, there is  substantial
evidence that  many of these objects are required in an underlying
theory  of strings and branes and as such  the
conditions found in this paper  provide the required generalisations
of the known BPS conditions. 
\par
This paper provides another example where $E_{11}$ predicts results that
are traditionally thought to be the result of supersymmetry. It would be
good to understand at a deeper level what is the connection between
$E_{11}$ and supersymmetry. 
\par
It was not clear from the outset that the  brane charges in all
dimensions should assemble into one $E_{11}$ multiplet, but we now know
that this is the case and the multiplet in question is the $l_1$
multiplet. In this paper we have seen that  half BPS conditions 
for all the charges, that is, space-time momentum, particle, string,
etc,   assemble into simple
$E_{11}$ equations.  This can be interpreted as evidence of an underlying
$E_{11}$ symmetry. 
\par
The one part of doubled field theory [28] that is not contained in the  
non-linear realisation of $E_{11}\otimes_s l_1$ at lowest level, as first
proposed in [10], is the so called section condition, which is
essentially equation (32).  It has been found that in most
circumstances this just sets to zero  the dependence on one coordinates.
A generalisation of this condition to the type of generalised space-time
at lowest level   first proposed in [10],   was given in reference [30].
This is essentially the condition of equation (13),  but applied  to the
usual U duality groups
$E_n$, $n=4,5,6,7,8$. We see that this  condition is  really one of the 
half BPS conditions, but  one might expect that the theory will involve
more than just half BPS states. One could for example impose the more
general quarter BPS conditions of equation (41). A first principle,
although limited,  approach to the question of how to reduce the
generalised space-time to something we recognise was given in [31]. 
\par
We end with a speculative comment on higher derivative string corrections
and the automorphic forms that describe them.  It has been observed that 
the automorphic forms that appear in front of the $R^4$ and $D^4 R$ terms
in type II string theory can be derived from the eleven dimensional one
[32] and two [33] loop diagrams respectively. The sums over the integers
arising from the Kaluza-Klein and other half BPS modes.  In particular
the auotomorphic form for the $R^4$ term arises from a scalar-like one
loop box diagram at zero external momentum which in nine dimensions is of
the generic form 
$$
\sum_{n_1, n_2}\int d^9 k {1\over( k^2 + M(n_1, n_2)^2)^4}
\eqno(42)$$
where $M(n_1, n_2)^2$ is the mass of the Klauza-Klein modes as it appears
in the half BPS condition. For lower dimensions one should also impose
the half BPS conditions of equation (21). There is now overwhelming
evidence that this automorphic form correctly accounts for all the $R^4$
corrections. 
\par
We note that $L^{-2}$ looks
like a propagator with a  mass that is one of the BPS states; it is also
$E_{11}$ invariant. If we were to take into account all  the BPS states
that should occur in the underlying theory, then we would expect to find
a coefficient for the $R^4$ term of the generic form 
$$
\sum {1\over (l^2)^4} 
\eqno(43)$$
The sum is to be interpreted as being  over the half BPS charges that
are active and this will depend on the dimension one is in. The sum will
include an integral over the  space-time momentum in the uncompactified
space-time and where appropriate  the Kaluza Klein modes (internal
momentum components), the winding modes (components of the  rank two
central charge), etc. If we include only low levels then the object in
equation (43)   will lead to the automorphic forms that arise in
equation (42). However, at higher levels one finds  terms  arising from
the fact that one has an $E_{11}$ automorphic form. A method of
constructing  automorphic forms for any group was given in [22,35] and
although this was applied to find the correct automorphic forms for the
usual finite U duality groups it can also be applied to
$E_{11}$, so leading also to the form given generically in equation
(43). We note that one must also by hand implement the half BPS conditions
of equation (13) on the automorphic form. 
\par
The $R^4$ part of the correction together with other terms constructed
from bosonic fields can hopefully  be constructed from
$E_{11}$ Cartan forms in such a way so as to be $E_{11}$ invariant. This
is to be expected as terms related by supersymmetry should have the same
automorphic form.  Thus  the effective action  should be composed of
different  $E_{11}$ invariant building blocks. From the $E_{11}$
perspective the different theories arise by taking different
decompositions of
$E_{11}$ and as a result they all have a common origin. An $E_{11}$
formulation of the higher string corrections would encode this common
origin. This is consistent with the fact that the BPS masses that appear
in equation (42) are the same regardless of whether they are computed
from the IIB, IIA or M theory perspectives [36]. Further ideas
concerning how higher derivative string corrections can admit an
$E_{11}$ formulation can be found in [37]. 

%%%%%%%%%%%%%%%%%%%%%%%%%%%%%%%%%%%%%%%%%%%%%%%%%%%%%%%%%%%%%

\medskip
{\bf {Acknowledgment}}
\medskip 
This work reported in this paper began with a number of discussions
between the author, David Berman and Malcolm Perry at Cooks Branch
Conservancy, Texas; I thank them for their help at an early stage of this
work. Notes containing the results in this paper were circulated to a
small set of people in Cambridge, England  in the summer of 2011.  I wish
to thank the STFC for support from grant number ST/J002798/1 and the
George Mitchell Foundation  for funding  for the visit to Cooks Branch
Conservancy.

%%%%%%%%%%%%%%%%%%%%%%%%%%%%%%%%%%%%%%%%%%%%%%%%%%%%%%%%%%%%%%
\medskip
{\bf {Appendix The $E_{11} \otimes_sl_1$ algebra}}
\medskip 
This appendix is designed to equip the reader with the $E_{11}$ material
required to understand this paper. Rather than explain the theory behind
Kac-Moody algebras we will present the required results. We first  give
the
$E_{11}$ algebra in the decomposition appropriate to eleven dimensions,
that is, we decompose  the $E_{11}$ algebra into representations   of
$A_{10}$, or SL(11), representations [4,10]. This algebra is found by
deleting node eleven in the Dynkin Diagram.  
The way one constructs the $E_{11}$  algebra, from the definition of
$E_{11}$ as a Kac-Moody algebra,  in terms of representations of SL(11) is
discussed, for example,  in [21]. For the calculation in this paper one
does not need to understand all the subtleties of this construction and
the parts of the algebra that are needed are given below. The generators
can be classified according to a level which is associated with the
decomposition associated with the deletion of  node eleven.  At level
zero we have the algebra  GL(11) with the generators
$K^a{}_b,\ a,b =1,\ldots 11$ and at level one and minus one the rank
three generators $R^{abc}$ and 
$R_{abc}$ respectively.  The generators at level two and minus two are
$R^{a_1\ldots a_6}$
 and $R_{a_1\ldots a_6}$ respectively, while those at levels three and
minus three are $R^{a_1\ldots a_8,b}$
 and $R_{a_1\ldots a_8,b}$ respectively. The level is just the number of
upper minus lower indices divided by three. For a discussion giving the
more abstract definition of level which relates it to the deletion of
node eleven see for example reference [21]. 
\par
The $E_{11}$ algebra at   levels zero and up  three is
given by [4,10]
$$
[K^a{}_b,K^c{}_d]=\delta _b^c K^a{}_d - \delta _d^a K^c{}_b, Ê
\eqno(A.1)$$
$$Ê [K^a{}_b, R^{c_1\ldots c_6}]=Ê
\delta _b^{c_1}R^{ac_2\ldots c_6}+\dots, \ Ê
Ê[K^a{}_b, R^{c_1\ldots c_3}]= \delta _b^{c_1}R^{a c_2 c_3}+\dots,
\eqno(A.2)$$
$$ [ K^a{}_b,Ê R^{c_1\ldots c_8, d} ]=Ê
(\delta ^{c_1}_b R^{a c_2\ldots c_8, d} +\cdots) + \delta _b^d
R^{c_1\ldots c_8, a} .
\eqno(A.3)$$
and 
$$[ R^{c_1\ldots c_3}, R^{c_4\ldots c_6}]= 2 R^{c_1\ldots c_6},\quad 
[R^{a_1\ldots a_6}, R^{b_1\ldots b_3}]
= 3Ê R^{a_1\ldots a_6 [b_1 b_2,b_3]},Ê
\eqno(A.4)$$
where $+\ldots $ means the appropriate anti-symmetrisation.Ê

The $E_{11}$ level zero  and negative level generators up to level minus
three obey the relationsÊ
$$
[K^a{}_b, R_{c_1\ldots c_3}]= -\delta ^a_{c_1}R_{b c_2
c_3}-\dots,\ [K^a{}_b, R_{c_1\ldots c_6}]=Ê -\delta ^a_{c_1}R_{bc_2\ldots
c_6}-\dots,
\eqno(A.5)$$
$$ [ K^a{}_b,Ê R_{c_1\ldots c_8, d} ]=Ê
-(\delta ^a_{c_1} R_{b c_2\ldots c_8, d} +\cdots) - \delta ^a_d
R_{c_1\ldots c_8, b} .
\eqno(A.6)$$
$$[ R_{c_1\ldots c_3}, R_{c_4\ldots c_6}]= 2 R_{c_1\ldots c_6},\quadÊ
[R_{a_1\ldots a_6}, R_{b_1\ldots b_3}]
= 3Ê R_{a_1\ldots a_6 [b_1 b_2,b_3]},Ê
\eqno(A.7)$$
Finally, the commutation relations between the positive and negative
generators Ê are given byÊ

$$[ R^{a_1\ldots a_3}, R_{b_1\ldots b_3}]= 18 \delta^{[a_1a_2}_{[b_1b_2}
K^{a_3]}{}_{b_3]}-2\delta^{a_1a_2 a_3}_{b_1b_2 b_3} D,\ Ê
[ R_{b_1\ldots b_3}, R^{a_1\ldots a_6}]= {5!\over 2}
\delta^{[a_1a_2a_3}_{b_1b_2b_3}R^{a_4a_5a_6]}
$$
$$
[ R^{a_1\ldots a_6}, R_{b_1\ldots b_6}]= -5!.3.3
\delta^{[a_1\ldots a_5}_{[b_1\ldots b_5}
K^{a_6]}{}_{b_6]}+5!\delta^{a_1\ldotsÊ a_6}_{b_1\ldotsÊ b_6} D ,\quadÊ
$$
$$
[ R_{a_1\ldots a_3}, R^{b_1\ldots b_8,c}]= 8.7.2
( \delta_{[a_1a_2 a_3}^{[b_1b_2b_3} R^{b_4\ldots b_8] c}-
Ê\delta_{[a_1a_2 a_3}^{[b_1b_2 |c|} R^{b_3\ldots b_8]} )
$$
$$
[ R_{a_1\ldots a_6}, R^{b_1\ldots b_8,c}]= {7! .2\over 3}
( \delta_{[a_1\ldotsÊ a_6}^{[b_1\dots b_6} R^{b_7 b_8] c}-
Ê\delta_{[a_1\ldotsÊ a_6}^{c[b_1\ldots b_5 } R^{b_6b_7 b_8]})
\eqno(A.8)$$
where $D=\sum_b K^b{}_b$, $\delta^{a_1a_2}_{b_1b_2}=
{1\over
2}(\delta^{a_1}_{b_1}\delta^{a_2}_{b_2}-
\delta^{a_2}_{b_1}\delta^{a_1}_{b_2})=
\delta^{[a_1}_{b_1}\delta^{a_2]}_{b_2}$ with similar formulae whenÊ
more indices are involved.Ê
\par
We also need the fundamental representation of $E_{11}$  associated with
node one, denoted by $l_1$. By definition this is the representation with
highest weight
$\Lambda_1$ which obeys 
$(\Lambda_1, \alpha_{  a})=\delta _{a,1}, \  a=1,2\ldots ,11$ 
where $\alpha_{ a}$ are the simple roots of $E_{11}$. In the
decomposition to Sl(11), corresponding to the deletion of node eleven, 
one finds that the $l_1$ representation  contains  the objects $P_a$, 
$Z^{ab}$ and
$Z^{a_1\ldots a_5} , a,b, a_1, \ldots =1,\ldots , 11$ corresponding to
levels zero, one and two respectively. We have taken the first object,
i.e.
$P_a$,  to have level zero by choice. Taking these to be generators
belong to  a semi-direct product algebra with those of $E_{11}$,   denoted
by $E_{11}\otimes _s l_1$,  their commutation relations with the level one
generators of $E_{11}$ are given by [10]
$$
[R^{a_1a_2a_3}, P_b]= 3 \delta^{[a_1}_b Z^{a_2a_3]}, \ Ê
[R^{a_1a_2a_3}, Z^{b_1b_2} ]= Z^{a_1a_2a_3 b_1b_2},\ Ê
$$
$$[R^{a_1a_2a_3}, Z^{b_1\ldots b_5} ]=Z^{b_1\ldots b_5[a_1a_2,a_3]}+
Z^{b_1\ldots b_5 a_1a_2 a_3}
\eqno(A.9)$$ 
These equations define the normalisation 
of the generators of the $l_1$ representation. The commutators of the
generators of the $l_1$ representation with those of GL(11) are given by 
$$
Ê[K^a{}_b, P_c]= -\delta _c^a P_b +{1\over
2}\delta _b^a P_c,\ Ê [K^a{}_b, Z^{c_1c_2} ]= 2\delta_b^{[c_1} Z^{|a|c_2]}
+{1\over 2}\delta _b^a Z^{c_1c_2},
$$
$$
[K^a{}_b, Z^{c_1\ldots c_5} ]= 5\delta_b^{[c_1} Z^{|a|c_2\ldots c_5]}
+{1\over 2}\delta _b^a Z^{c_1\ldots c_5}
\eqno(A.10)$$
The commutation relations with the level two generators of $E_{11}$ are
given by 
$$
[R^{a_1\dots a_6}, P_b]= -3 \delta^{[a_1}_b Z^{\ldots a_6]}, \ 
[R^{a_1\dots a_6}, Z^{b_1b_2} ]= Z^{b_1b_2[a_1\ldots a_5,a_6]},\ Ê
\eqno(A.11)$$
TheÊ commutators with the level $-1$ negative root generatorsÊare given
by 
$$
[R_{a_1a_2a_3}, P_b ]= 0,\ Ê
[R_{a_1a_2a_3}, Z^{b_1b_2} ]= 6\delta^{b_1b_2 }_{[a_1a_2} P_{a_3 ]},\ Ê
[R_{a_1a_2a_3}, Z^{b_1\ldots b_5} ]= {5!\over 2} \delta^{[ b_1b_2b_3
}_{a_1a_2a_3} Z^{b_4b_5]}
\eqno(A.12)$$

%%%%%%%%%%%%%%%%%%%%%%%%%%%%%%%%%%%%%%%%%%%%%%%%%%%%%%%%%%%%%%%%%%%%%%%%%%%
\medskip 
{\bf References} 
\medskip
\item{[1]} E. Witten and D. Olive, {\it Supersymmetry algebras that
include topoloogical charges}, Phys. Lett. {\bf 78B} (1978) 97.
\item{[2]} J. Azcarraga, J. Gauntlett, J. Izquierdo and P. Townsend,Ê
{\it Topological Extensions of the Supersymmetry Algebra for Extended
Objects}, Phys. Rev. Lett. {\bf 63, no 22} (1989) 2443.Ê
\item{[3]} S. Ferrra and J Maldecena, {\it Branes, central charges and
U-duality invariant BPS conditions}, arXiv:hep-th/9706097; B. Pioline and 
E Kiritsis, {\it U-duality and D-brane Combinatorics}, Phys.Lett. {\bf
B418} (1998) 61, arXiv:hep-th/9710078. 
\item{[4]} P. West, {\it $E_{11}$ and M Theory}, Class. Quant.  
Grav.  {\bf 18}
(2001) 4443, {\tt arXiv:hep-th/ 0104081}; 
\item{[5]} I. Schnakenburg and  P. West, {\it Kac-Moody   
symmetries of
IIB supergravity}, Phys. Lett. {\bf B517} (2001) 421, {\tt  
arXiv:hep-th/0107181}.
\item{[6]}  F. ÊRiccioni and P. West, {\it
The $E_{11}$ origin of all maximal supergravities}, ÊJHEP {\bf 0707}
(2007) 063; ÊarXiv:0705.0752.
\item{[7]} ÊF. Riccioni and P. West, {\it E(11)-extended spacetime
and gauged supergravities},
JHEP {\bf 0802} (2008) 039, ÊarXiv:0712.1795
\item{[8]} ÊF. Riccioni and P. West,
Ê{\it Local E(11)}, JHEP {\bf 0904} (2009) 051, arXiv:hep-th/0902.4678.
\item{[9]} F. ÊRiccioni, ÊD. ÊSteele and P. West, {\it The E(11)
origin of all maximal supergravities - the hierarchy of field-strengths}
ÊÊJHEP {\bf 0909} (2009) 095, arXiv:0906.1177. 
\item{[10]} P. West, {\it $E_{11}$, SL(32) and Central Charges},
Phys. Lett. {\bf B 575} (2003) 333-342, {\tt hep-th/0307098}
\item{[11]}  A. Kleinschmidt and P. West, {\it  Representations of G+++
and the role of space-time},  JHEP 0402 (2004) 033,  hep-th/0312247.
\item{[12]} P. West,  {\it $E_{11}$ origin of Brane charges and U-duality
multiplets}, JHEP 0408 (2004) 052, hep-th/0406150. 
\item{[13]} P. Cook and P. West, {\it Charge multiplets and masses
for E(11)}, ÊJHEP {\bf 11} (2008) 091, arXiv:0805.4451.
\item {[14]} P. West, {\it The IIA, IIB and eleven dimensional theories 
and their common
$E_{11}$ origin}, Nucl. Phys. B693 (2004) 76-102, hep-th/0402140. 
\item{[15]} S. Elitzur, A. Giveon, D. Kutasov and E.Ê Rabinovici,Ê {\it
Algebraic aspects of matrix theory on $T^d$ }, {\tt Ê
arXiv:hep-th/9707217}.
\item{[16]}Ê N. Obers,Ê B. Pioline and E.Ê Rabinovici, {\it M-theory and
U-duality on $T^d$ with gauge backgrounds}, {\tt hep-th/9712084}
\item{[17]} N. Obers and B. Pioline,~ {\it U-duality and Ê
M-theory, an 
algebraic approach}~, {\tt hep-th/9812139}.
\item{[18]} B. de Wit and H. Nicolai, {\it Hidden symmetries, central
charges and all that}, hep-th/0011239.Ê
\item{[19]} N. Obers and B. Pioline,~ {\it U-duality and Ê
M-theory}, {\tt arXiv:hep-th/9809039}.
\item{[20]} P. West, {\it Brane dynamics, central charges and
$E_{11}$}, hep-th/0412336. 
\item{[21]} P. West,{\it  Introduction to Strings and Branes}, Cambridge
University Press, June 2012. 
\item{[22]} P. West and N. Lambert, {\it Duality Groups, Automorphic Forms
and Higher Derivative Corrections},  Phys.Rev. D75 (2007) 066002, 
hep-th/0611318
\item{[23]} T. Damour, M. Henneaux and H. Nicolai,
  {\it E(10) and a 'small tension expansion' of M theory},
  Phys. Rev. Lett.  {\bf 89} (2002) 221601, arXiv:hep-th/0207267.
\item{[24]} P. West, {\it Very Extended $E_8$ and $A_8$ at low
levels, Gravity and Supergravity}, Class.Quant.Grav. {\bf 20} (2003)
2393, hep-th/0212291.
\item{[25]}T. Nutma, SimpLie, a simple program for Lie algebras,
https://code.google.com/p/simplie/. 
\item{[26]} D.  Berman, H. Godazgar, M. Perry and  P.  West, 
{\it Duality Invariant Actions and Generalised Geometry}, 
arXiv:1111.0459. 
\item{[27]}  P. West, {\it E11, generalised space-time and IIA string
theory},  Phys.Lett.B696 (2011) 403-409,   arXiv:1009.2624.
\item{[28]} O. Hohm, C. Hull and B. Zwiebach, {\it Generalised metric
formulation of double field theory},  hep-th/1006.4823; 
\item{[29]}   A. Rocen and P. West,  {\it E11, generalised space-time and
IIA string theory;  the R-R sector},  arXiv:1012.2744.
\item{[30]}  A. Coimbra, C. Strickland-Constable and  D.  Waldram, 
 {\it $E_{d(d)} \times {R}^+$ Generalised Geometry,
Connections and M theory}, arXiv:1112.3989.
\item{[31]} P. West, {\it Generalised space-time and duality},
hep-th/1006.0893. 
\item{[32]} M.B. Green, M. Gutperle and  P. Vanhove, {\it  One loop in
eleven dimensions}, Phys.Lett. {\bf B409} (1997) 177,
arXiv:hep-th/9706175. 
\item{[33]}J. Russo and A.A. Tseytlin, {\it One-loop four-graviton
amplitude in eleven-dimensional supergravity}, Nucl.Phys. {\bf B508}
(1997) 245,  arXiv:hep-th/9707134. 
\item{[34]} Michael B. Green and  Hwang-h. Kwon, Pierre Vanhove, {\it Two
loops in eleven dimensions}, Phys.Rev. {\bf D61}(2000) 104010,
arXiv:hep-th/9910055.  
\item{[35]} N. Lambert and P. West, {\it Perturbation Theory From Automorphic Forms}, 
JHEP 1005 (2010) 098,Ê arXiv:1001.3284.Ê
\item{[36]} J. Schwarz, {\it An SL(2,Z) Multiplet of Type IIB
Superstrings}, Phys.Lett. B360 (1995) 13; Erratum-ibid. B364 (1995) 252, 
arXiv:hep-th/9508143; 
B, de Wit and  D. Lust, {\it BPS Amplitudes, Helicity Supertraces and
Membranes in M-Theory}  
Journal-ref: Phys.Lett.{\bf B477} (2000) 299, arXiv:hep-th/9912225
\item{[37]}F. Gubay and  P.  West, {\it Higher derivative type
II string effective actions, automorphic forms and E11}, arXiv:1111.0464.

\end